\documentclass[twocolumn]{article}

\usepackage{amsmath}
\usepackage{graphicx}
\usepackage[margin=0.7in]{geometry}
\usepackage{siunitx}
\usepackage{url}
\usepackage{cite}
 
\usepackage{esint}
\usepackage{xcolor}
\usepackage{lineno}
\usepackage{gensymb}
\usepackage{subfigure}

\usepackage{authblk}
\usepackage{caption}
\usepackage{blindtext}
\makeatletter
\renewcommand*\env@matrix[1][\arraystretch]{%
  \edef\arraystretch{#1}%
  \hskip -\arraycolsep
  \let\@ifnextchar\new@ifnextchar
  \array{*\c@MaxMatrixCols c}}
\makeatother

\begin{document}
\title{The role of the Silberstein/Thomas/Wigner-rotation in the rod and slit paradox}
\author{
  Mads Vestergaard Schmidt\thanks{e-mail: thesparetimephysicist@gmail.com} \
  Erich Schoedl
}

\maketitle

\begin{abstract}
\sl{The rod and slit paradox, as first proposed by R. Shaw, is revisited. In this paradox, a rod of rest length $l_0$ moves parallel to the horizontal x-axis. Simultaneously, a thin sheet parallel to the horizontal xz-plane with a rod-shaped slit of rest length $d < l_0$ moves along the vertical y-axis. The rod passes through the slit due to relativistic effects. Earlier papers do not link the thought experiment to the Silberstein/Thomas/Wigner-rotation (STW-rotation), which is essential to its outcome. In this paper the role of the STW-rotation is discussed, and additional clarifying figures are provided. Some misleading aspects of earlier treatments are pointed out and corrected.}  
\end{abstract}

\section{Introduction}

Apparent length contraction paradoxes in which a rod with contracted length $l = \frac{l_0}{\gamma}$ passes through a parallel sheet with a slit of length $d<l_0$, exist in several versions \cite{Rindler, Martins, Gron, Van Lintel}. Here we will be concerned with the case where the rod moves horizontally and the parallel sheet moves vertically as first proposed by R. Shaw \cite{Shaw} (1962). No accelerations are present in this experiment as the rod approaches the slit. If the rod and sheet are given appropriate trajectories and velocities the contracted rod passes through the slit. The apparent paradox arises when the situation is viewed in the rest frame of the rod where $l > d$, and it is resolved by showing that rotational effects occur between the rod and the slit, allowing the longer rod to clear the shorter slit. In Special Relativity there are two distinct rotational effects. One is due to length contraction and the other, the Silberstein/Thomas/Wigner-rotation (STW-rotation), occurs relative to all reference frames in non-collinear motion, as a boost/transformation is performed.
 
The focus of this paper is to show how the STW-rotation is essential to the resolution of the rod and slit paradox, and to clarify some misleading aspects of earlier treatments. In particular we will discuss the solutions presented by E. Marx \cite{Marx} (1967) and by Iyer and Prahbu \cite{Prabhu} (2006), neither of which addresses the role of the STW-rotation.

In Marx’s paper a minor misrepresentation of the trajectories in his figures, combined with a distortion of the dimension of the slit, blurs the fact that two of the displayed reference systems are STW-rotated with respect to each other. Marx never claimed that his figures are drawn to scale, but nevertheless this feature can be confusing to the reader and deserves to be clarified. Iyer and Prabhu (2006) analyzed a direct transformation between the rest frames of the rod and the slit and concluded that the passing of the rod through the slit is unaffected by relativistic kinematics. In this treatment, the rod and slit proportions and orientations are not clearly defined in the initial reference frame defined by Shaw, so the analysis ignores the non-collinear velocities that determine the STW-rotation, and decide the outcome of the experiment. Iyer and Prabhu argue that that their conclusion applies to Shaw’s version as well as Marx’s second example versions of the paradox, while we will argue that it does not. It should be noted that Ferraro (2007) \cite{Ferraro_book} correctly links the solution of the paradox to the STW-rotation. This paper will add to Ferraro’s treatment by discussing the issues in previous work, as well as providing additional figures describing the STW-rotation for this paradox.

\section{Discussion of earlier treatments} \label{part1}
\label{Earlier treatments}

\begin{figure}[h!]
    \centering
    \fbox{\subfigure[]{\includegraphics[width=0.30\textwidth]{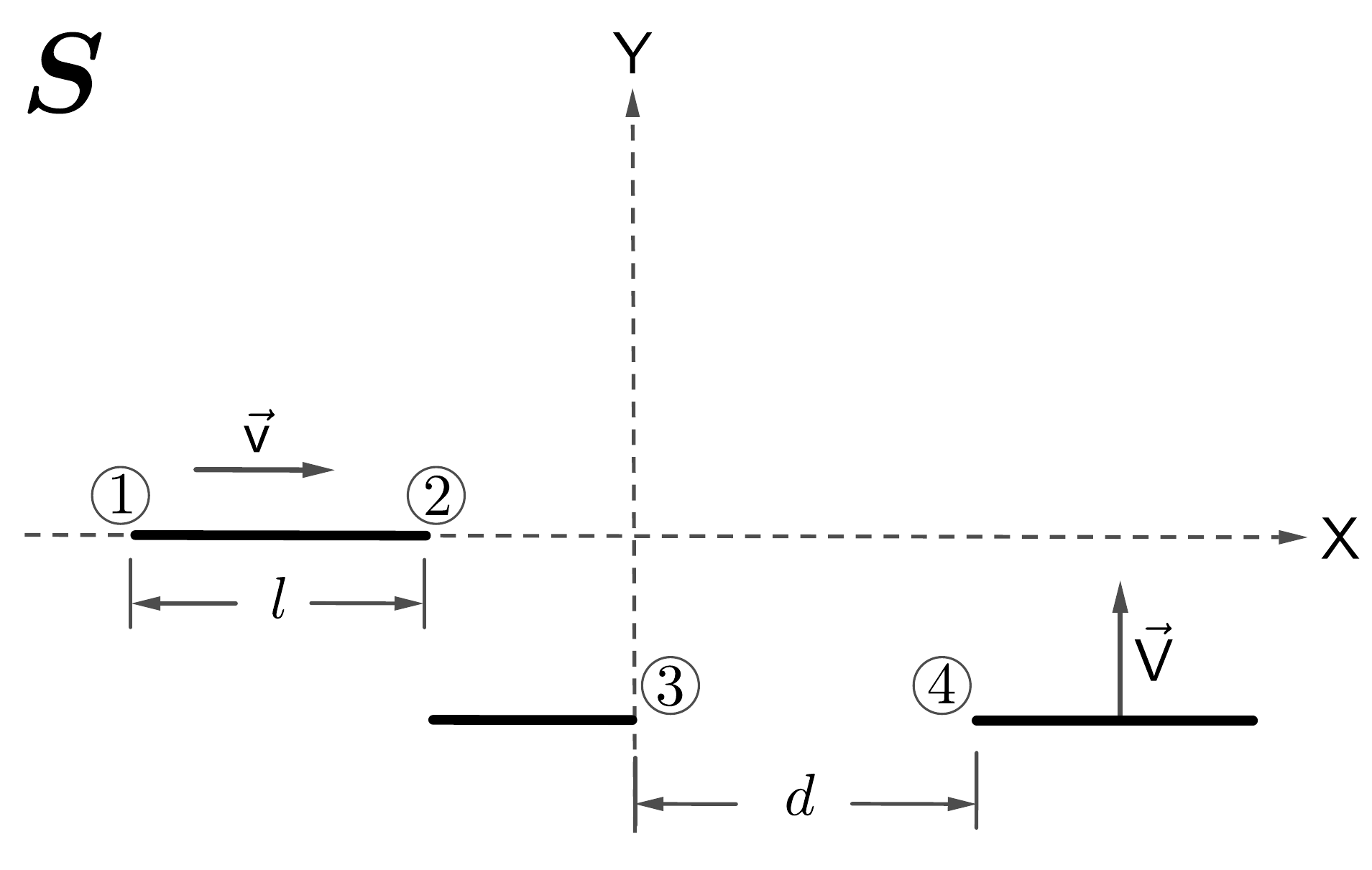}}}
       
   \fbox{ \subfigure[]{\includegraphics[width=0.30\textwidth]{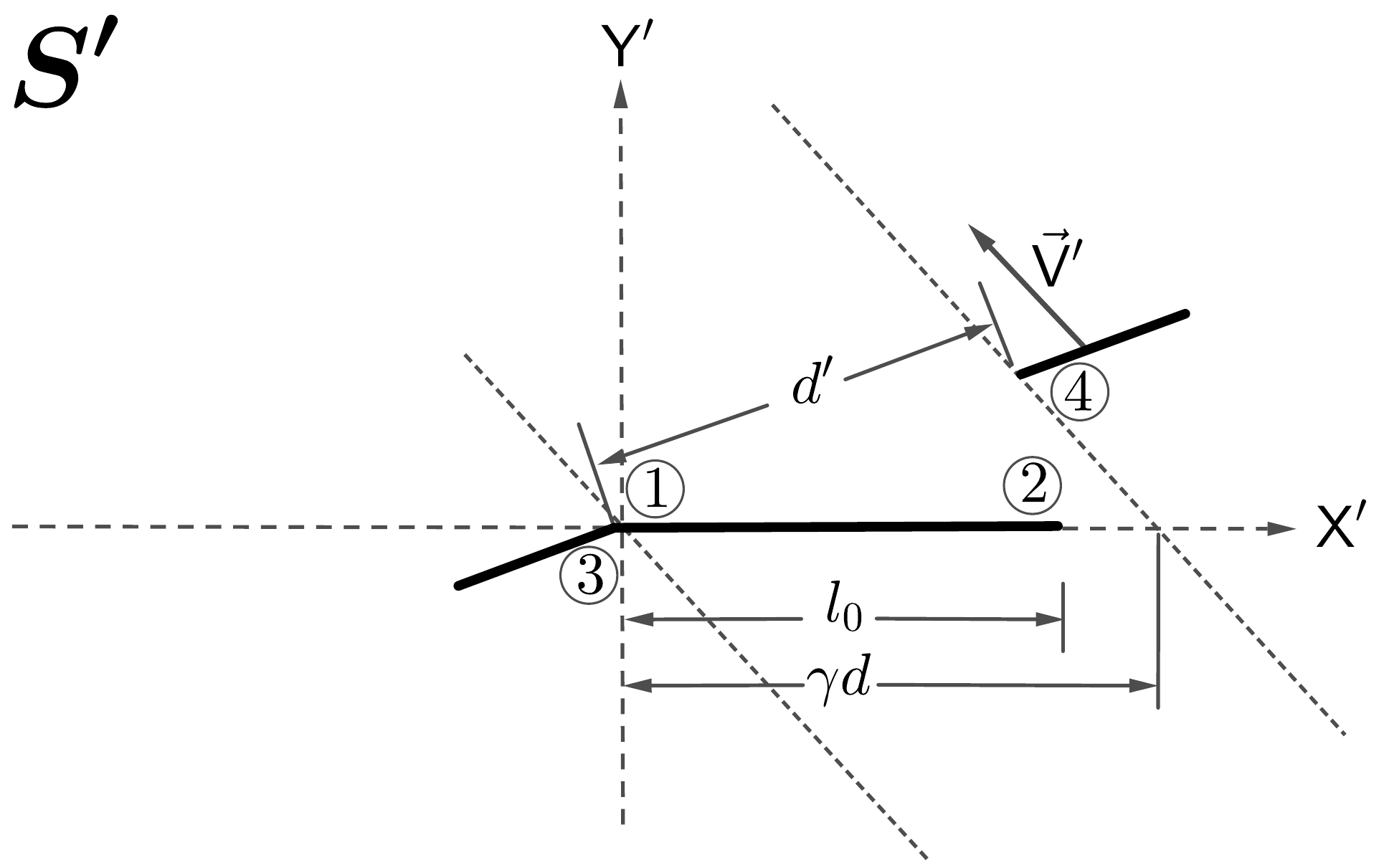}}}
    
   \fbox{ \subfigure[]{\includegraphics[width=0.30\textwidth]{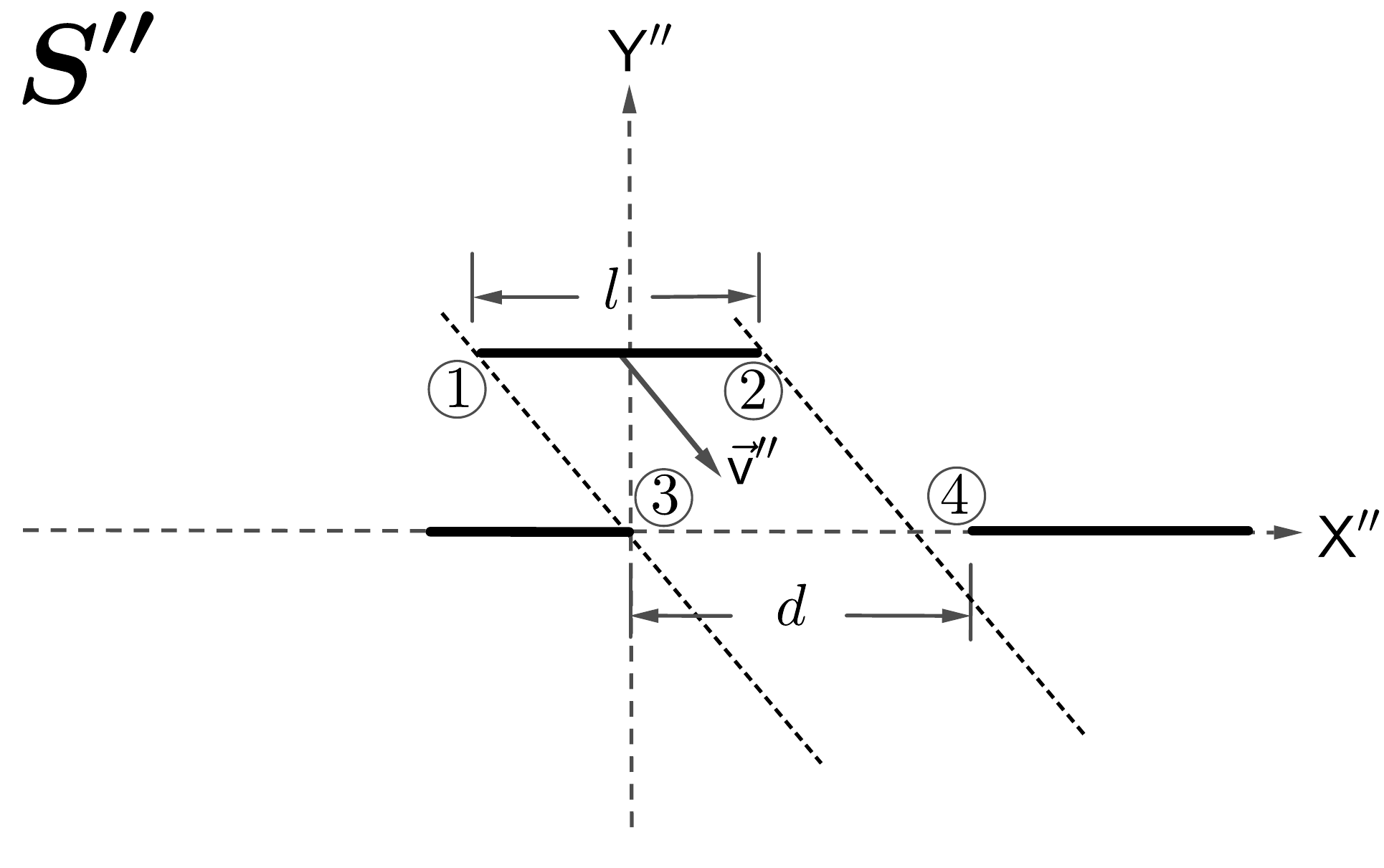}}}

    \caption{\sl Recreation of Marx’s Figure 2 showing his second experiment where (a) the rod moves along the positive X axis and the sheet moves at a right angle along the positive Y axis as seen by an observer at rest on the ground, (b) the same experiment observed from the rest frame of the rod, and (c) observed from the rest frame of the sheet.}
    \label{fig1}
\end{figure}

In the "rod and slit" paradox (also referred to as the "bar and ring" or "rod and slot" paradox), a rod of contracted length $l$ moves along the horizontal x-axis at velocity $\vec{v}$. At the same time a sheet, parallel to the horizontal xz-plane, with a slit inside moves along the vertical y-axis at velocity $\vec{V}$. The situation is illustrated in Fig. \ref{fig1}a. If the motions of the two objects are carefully coordinated, and v is sufficiently large, it is possible for the contracted rod to pass through the slit even if the proper length of the rod $l_0$ is longer than the proper length of the slit $d$ and would collide in non-relativistic mechanics. The apparent paradox arises when the situation is viewed in a commoving frame with the rod, or "the rest frame of the rod", $S'$. Because $l_0$ is larger than $d$ in this frame, one might assume that the rod would not pass through the slit. The apparent paradox is easily resolved by transforming the coordinates from $S$ to $S'$, as shown in Fig.  \ref{fig1}b. This transformation reveals that the sheet is rotated with respect to the rod so that the longer rod clears the slit. In the treatment by Marx (1967) the situation is also displayed in the rest frame of the sheet, $S''$, as seen in Fig. 1c. The difference in simultaneity between frames is responsible for the rotation of the sheet in frame $S'$. Since point 4 is ahead in time of point 3, compared to frame $S$, its vertical position is also ahead and the sheet becomes tilted. It may be confusing to see the rod is not conversely rotated in frame $S''$. The reason for this is that the extension of the rod’s length $l$ is orthogonal to the direction of the boost from frame $S$ to $S''$, and so the relativity of simultaneity causes no rotation to the observed orientation of the rod. The rotations in the relative frames will be expounded on in more detail in section 3, and calculations of the coordinates in the three frames are presented in Appendix A.

Fig. \ref{fig1} is a recreation of Figure 2, and in Fig. \ref{fig1}b and \ref{fig1}c we see that the velocities, $\vec{V'}$ and $\vec{v''}$, are approximately anti-parallel. The velocities can be obtained by using the relativistic velocity addition formula \cite{Ferraro_book_V}. Using  $\gamma_v = (1-(v/c)^2)^{-1/2}$, and $\gamma_{\scriptscriptstyle V} = (1- (V/c)^2)^{-1/2}$ we get

\begin{align}
\vec{V'} = \begin{bmatrix}[2]
           \frac{V_x - v}{1-v V_x/c^2} \\
          \frac{V_y \sqrt{1-v^2/c^2}}{1-v V_x/c^2} \\
          \end{bmatrix} =
          \begin{bmatrix}[1.5]
           -v \\
          V/\gamma_v \\
          \end{bmatrix}
\label{Velocity1}
\end{align}

\begin{align}
\vec{v''} =\begin{bmatrix}[2] \frac{v_x \sqrt{1-V^2/c^2}}{1-V v_y/c^2} \\   \frac{v_y - V}{1-V v_y/c^2} \end{bmatrix} = \begin{bmatrix}[1.5] -v/\gamma_{\scriptscriptstyle V} \\ V   \end{bmatrix}
\label{Velocity2}
\end{align}

\begin{figure}[h!]
    \centering
    \fbox{\subfigure[]{\includegraphics[width=0.30\textwidth]{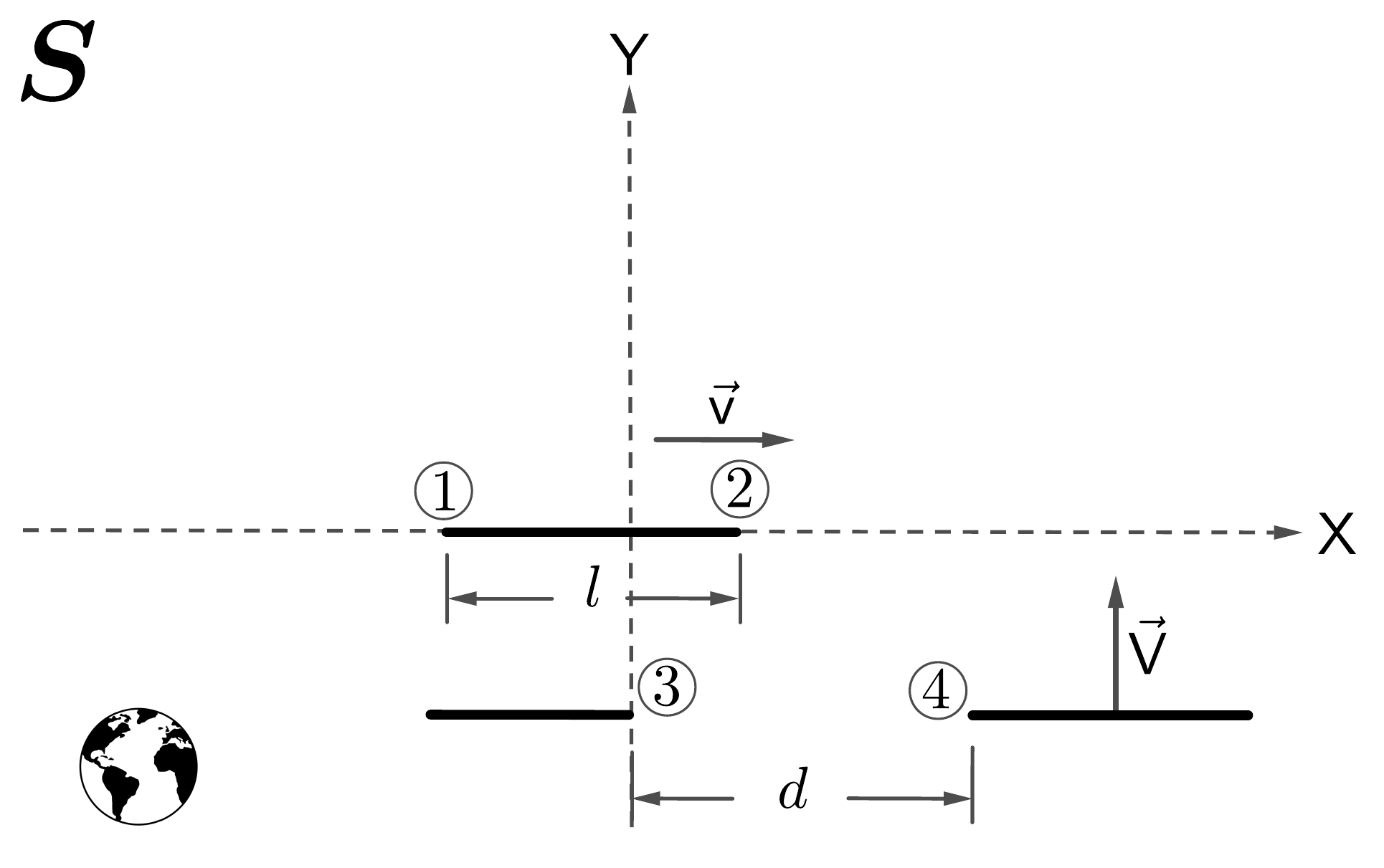}}}    
    
    \fbox{\subfigure[]{\includegraphics[width=0.30\textwidth]{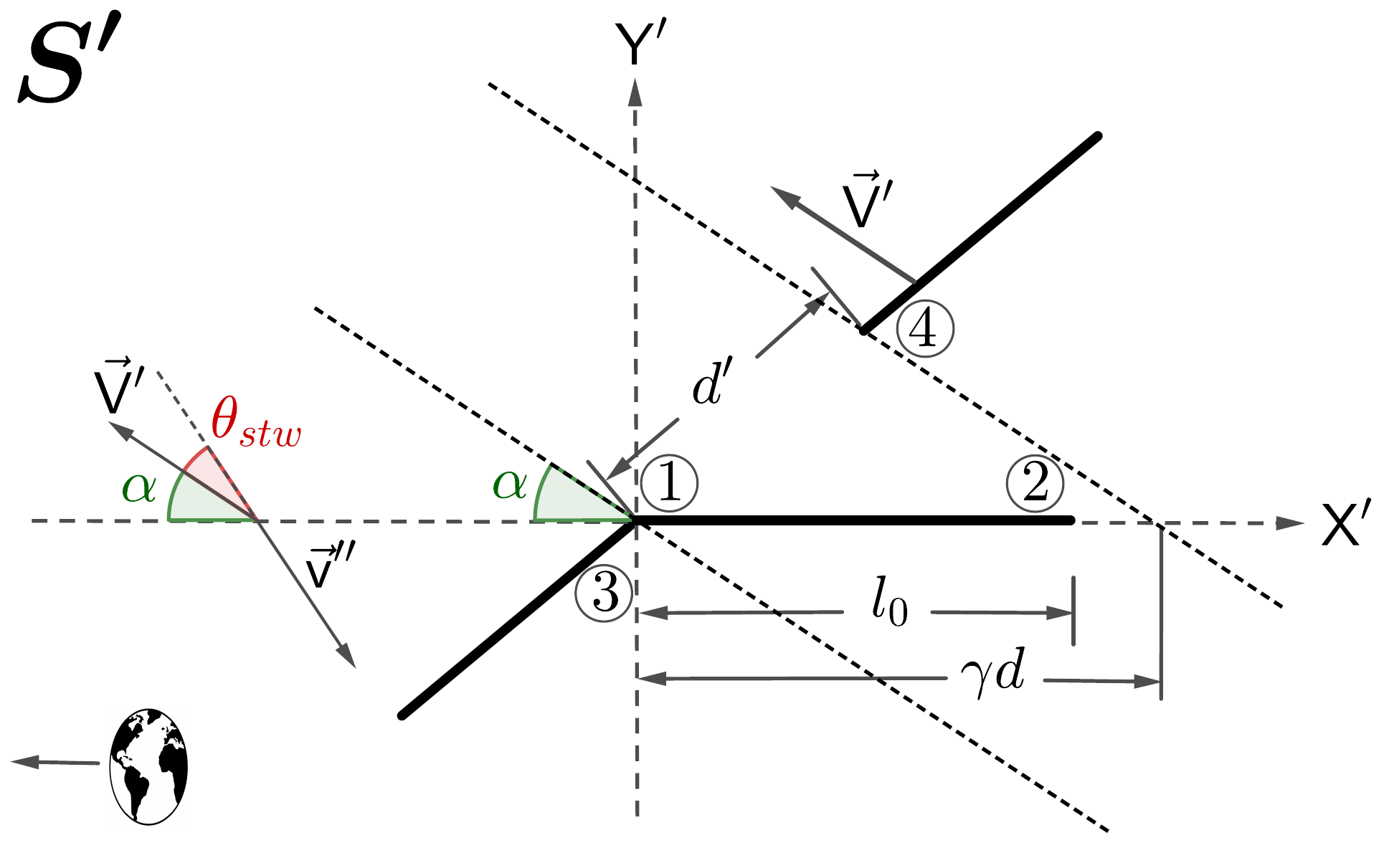}}}
       
   \fbox{ \subfigure[]{\includegraphics[width=0.30\textwidth]{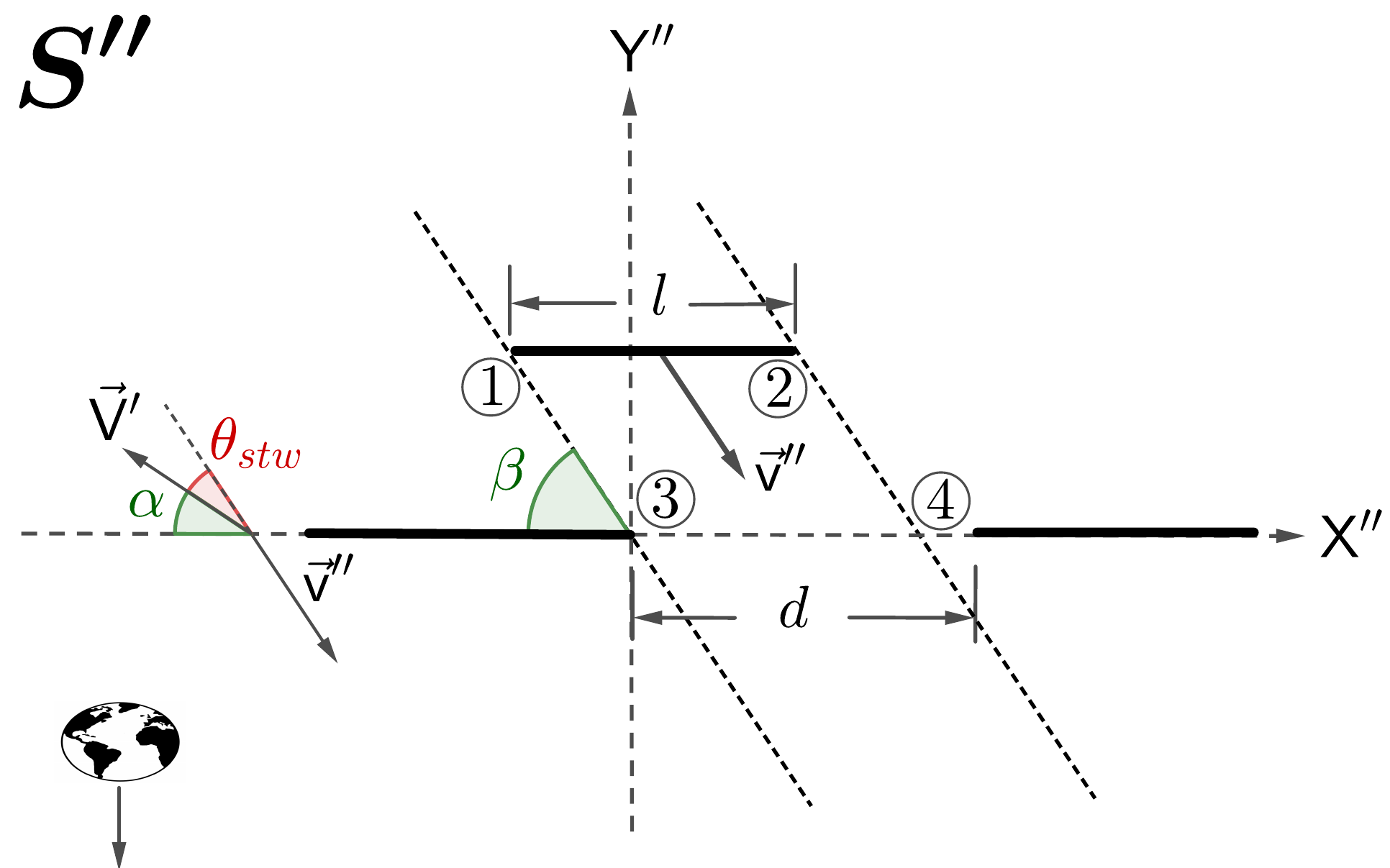}}}

    \caption{\sl Properly scaled representation of experiment defined in Fig. \ref{fig1} for $v = V$, $d = 0.785 \cdot l_0$, $\gamma_v = 1.5$, and the trajectory aligns through points 1 and 3 where (a) shows the initial rest frame with two non-collinear velocities, (b) shows the case from the rest frame of the rod, and (c) from the rest frame of the slit. The globe in the lower left corner shows the relative direction of motion of frame $S$.}
    \label{fig2}
\end{figure}

Eqs. \ref{Velocity1} and \ref{Velocity2} shows that $\vec{V'}$ and $\vec{v''}$ are not anti-parallel as Marx’s figure otherwise suggests. The reason for this is that the coordinate systems of $S'$ and $S''$ are in fact rotated with respect to each other. This rotation is known as the STW-rotation (or Wigner-rotation). The effect was discovered by L. Silberstein in 1914 \cite{Silberstein}, rediscovered by L. Thomas in 1921 \cite{Thomas}, and derived in a general form by E. Wigner in 1939 \cite{Wigner,Ferraro}. Relative STW-rotation between two frames of reference, occurs when two or more non-collinear Lorentz boosts are performed to get from one to the other. The rotation angle, $\theta_{stw}$, is the angle between the relative velocity vectors of the frames \cite{Visser}. In Marx’s treatment the frames $S'$ and $S''$ are related by two non-collinear Lorentz boosts, resulting in an STW-rotation angle of 

\begin{align}
\theta_{stw} = cos^{-1} \left( \frac{\vec{V'} \cdot \vec{v''}}{\parallel \vec{V'} \parallel \parallel \vec{v''} \parallel}\right) .
\label{Wigner}
\end{align}

Specific details of how the STW-rotation is derived can be found in Ferraro and Thibeault \cite{Ferraro} and Ferraro \cite{Ferraro_book_Wig}. Multiple methods of calculating the angle itself are discussed by O’Donnell and Visser \cite{Visser}.

Figure 2 shows a properly scaled version of Marx’s Figure 2 where we have chosen comparable parameters $v = V$, $d = 0.785 \cdot l_0$, and $\gamma_v = 1.5$. (The length of the slit, $d$, must be between $l_0/ \gamma_v$ and $l_0$ in order to be meaningful, and the arbitrary choice of $d=0.785 \cdot l_0$ is chosen to closely match the original figures by Marx.) In the special case where speeds $\mid v \mid = \mid V \mid$, and the velocities are orthogonal in the $S$ frame, we can use  Eqs. \ref{Velocity1} and \ref{Velocity2} to simplify the STW-rotation formula to $\theta_{stw} = cos^{-1}(2/(\gamma_v + 1/\gamma_v))$. This gives us the STW-angle between $\vec{V'}$ and $\vec{v''}$ of $22.6 \degree$. We clearly see the different trajectory angels given in Figs. \ref{fig2}b and \ref{fig2}c. It is also evident that the dimensions of the slit in Fig. \ref{fig1}b are distorted relative to Marx’s Figs. \ref{fig1}a and \ref{fig1}c. 
\\
\\

\begin{figure}[h!]
    \centering
    \includegraphics[width=0.30\textwidth]{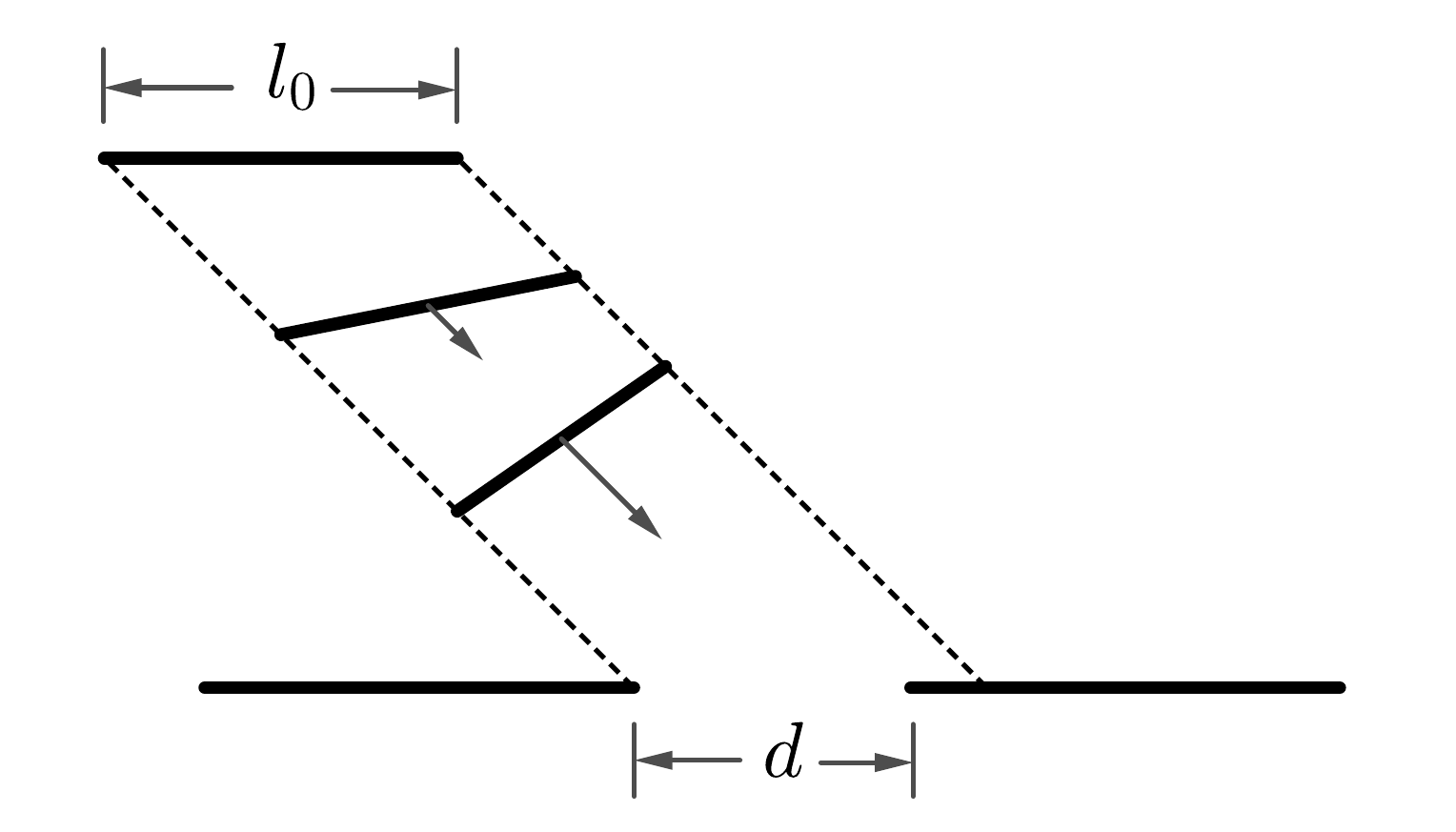} 
    \caption{\sl Recreation of Marx’s figure 1 for his first experiment showing the rod moving directly toward the slit with increased contraction for increased speeds causing an apparent rotation, but the paths defining the ends do not rotate or change for the single relativistic velocity case for any speed since there is no STW-rotation.}
    \label{fig3}
\end{figure}

In his paper, Marx discusses two different thought-experiments. Experiment II is the original paradox by Shaw that is described above. In experiment I, the sheet is kept stationary while the rod is given speed directly towards the slit. Experiment I is displayed in Fig. \ref{fig3}. The single boost does not give rise to any real or STW-rotations, and observed rotation is purely due to length contraction. Since the trajectories of the edges of the rod are independent of the speed, this “contraction” rotation has no effect on the outcome of the experiment.

For both thought-experiments an initial non-relativistic and a final relativistic state are defined. It is specified that in the non-relativistic state the rod and sheet are parallel with respect to each other, and we must assume that the relativistic state is reached by performing boosts parallel to the respective velocity vectors. With these conditions no rotational forces act on the objects as the boosts are performed. Here Marx's aim is to show that relativistic boosts are responsible for the passing of the rod in experiment II, and that length contraction is a real rather than an apparent effect.
\\
\\
In Iyer and Prabhu’s setup \cite{Prabhu} (2006), an initial state of the system is not defined, and it is not explained how the velocities of the rod and slit are increased to reach the final state of the system. Therefore, it is unclear how they distinguish between Marx’s experiments I and II. In their first example, case 1, they use a system that resembles $S''$, as displayed in Fig. \ref{fig2}c, and transform the coordinates to the rest frame of the rod by performing a direct transformation from $S''$ to $S'$. From this transformation the proper angles of the rod and the slit relative to the trajectory are obtained, and these serve as the basis of their conclusion. The logic goes: If the rod and slit are rotated sufficiently with respect to each other the rod will pass, and since the same condition can be established in Galilean kinematics, the outcome is independent of relativistic kinematics. The problem with this logic is that the initial condition that induces the rotation is not correctly established. In Galilean kinematics only classical rotation is possible, but in relativistic kinematics it is possible to create a real relative rotation without classically rotating the objects as per the originally proposed Shaw paradox. Marx's thought-experiment I teaches us that the relative contraction between two rest frames does not influence the outcome of the paradox, but with the two initial non-collinear velocities in case II a relative STW-rotation between the rod and the slit is induced. Iyer and Prabhu simply fail to recognize this relative rotation as a relativistic effect, and therefore reach the faulty conclusion that the outcome of the experiment by Shaw (or Marx's experiment II) is independent of relativistic kinematics. In section \ref{part2} we provide an analysis, similar to the one by Iyer and Prabhu, to clearly show how the STW-rotation is decisive for the outcome in the rest frame perspective.

\section{Additional analysis} \label{part2}

The STW-rotation can be a challenging concept to grasp intuitively, and compared to length contraction, time dilation, and the relativity of simultaneity, there is a lack of educational examples in the literature. While the rod and slit paradox was originally presented as a length contraction paradox, it is even more relevant as a way to exemplify the STW-rotation. We will now use the STW-rotation to discern whether or not the passing of the rod is determined by relativistic kinematics. 

Let us define $\alpha$ to be the proper angle of the rod, and $\beta$ to be the proper angle of the slit relative to the direction of motion. These can be seen in Fig. \ref{fig2}b and \ref{fig2}c. Because both the rod and slit are aligned with the x-axis in their respective rest frames, the difference in angle between $\alpha$ and $\beta$ is the angle between $\vec{V'}$ and $\vec{v''}$. As described by Eq. \ref{fig3}, this is the STW-angle, and we have $\beta = \alpha + \theta_{stw} $. By comparing the projection of the rod and slit onto the plane perpendicular to the direction of motion for either proposed case, one can determine if the rod will clear the slit. For the case studied here the rod will pass the slit if

\begin{equation}
\mid sin(\mid \alpha \mid + \theta_{stw}) \mid d > \mid sin( \mid \alpha \mid) \mid l_0.
\label{ExperimentCondition}
\end{equation}

In the case of non relativistic velocities we get $\theta_{stw}\approx 0$, and because $l_0>d$ the condition of Eq. \ref{ExperimentCondition} is not met. In the case of relativistic velocities $\theta_{stw} \neq 0$, and with the appropriate choice of parameters it is possible to get the longer rod to pass the shorter slit.
Since $l_0$ and $d$ are constants in the experiment, we can conclude that the outcome solely depends on $\theta_{stw}$, induced by the non-collinear velocities chosen relative to the frame $S$. Since the STW-rotation is a distinct relativistic effect with no classical equivalent, the outcome of the experiment is clearly determined by relativistic kinematics.

It should be noted that the condition for passing, $d > l_0/ \gamma$, can be defined in the $S$ frame so that length contraction is the decisive factor from this perspective, as pointed out by Shaw. This is because the axes in frame $S$ have no STW-rotation with respect to $S'$ and $S''$. 

\begin{figure}[h!]
    \centering
    \fbox{\subfigure[]{\includegraphics[width=0.30\textwidth]{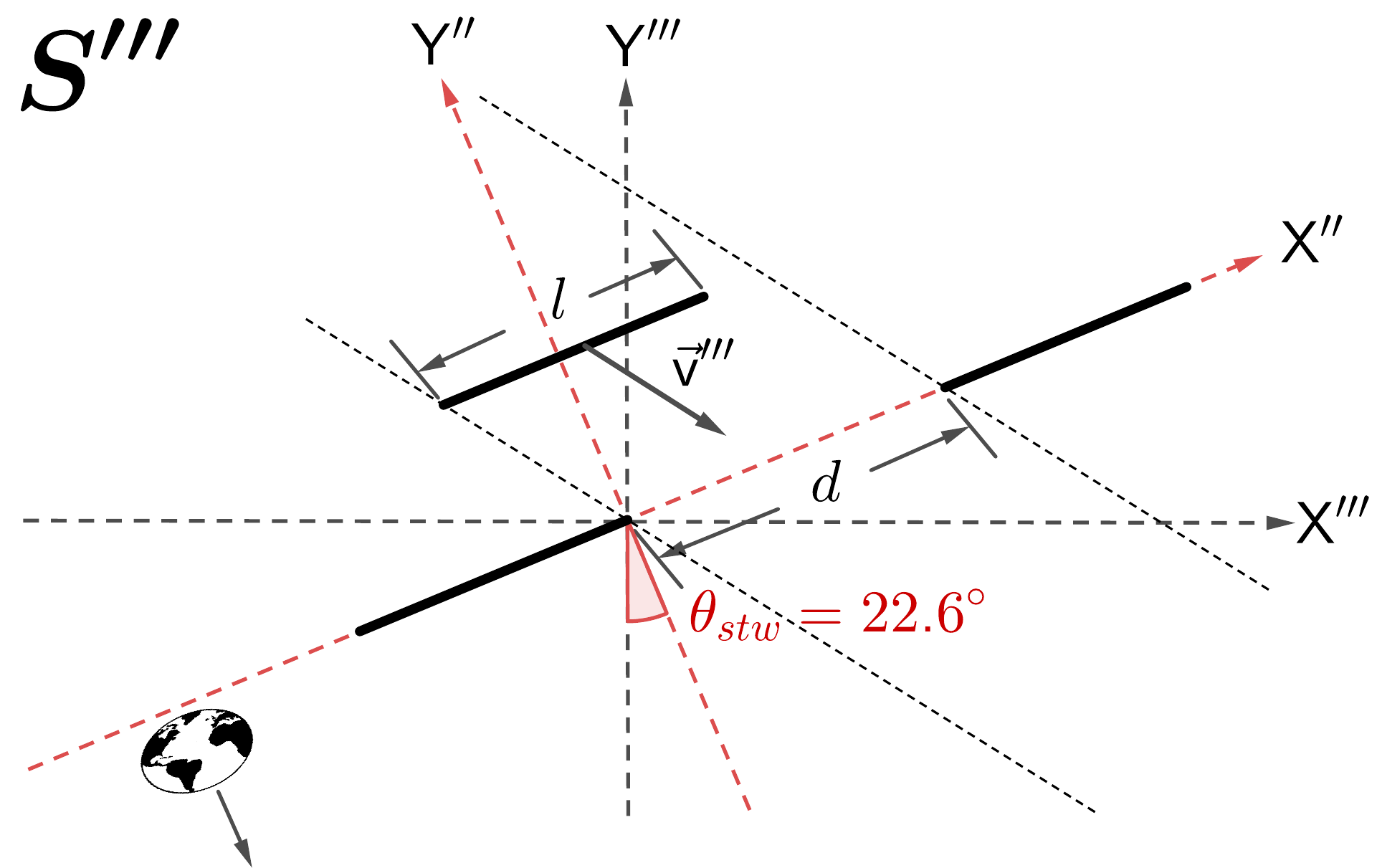}}}
       
   \fbox{ \subfigure[]{\includegraphics[width=0.30\textwidth]{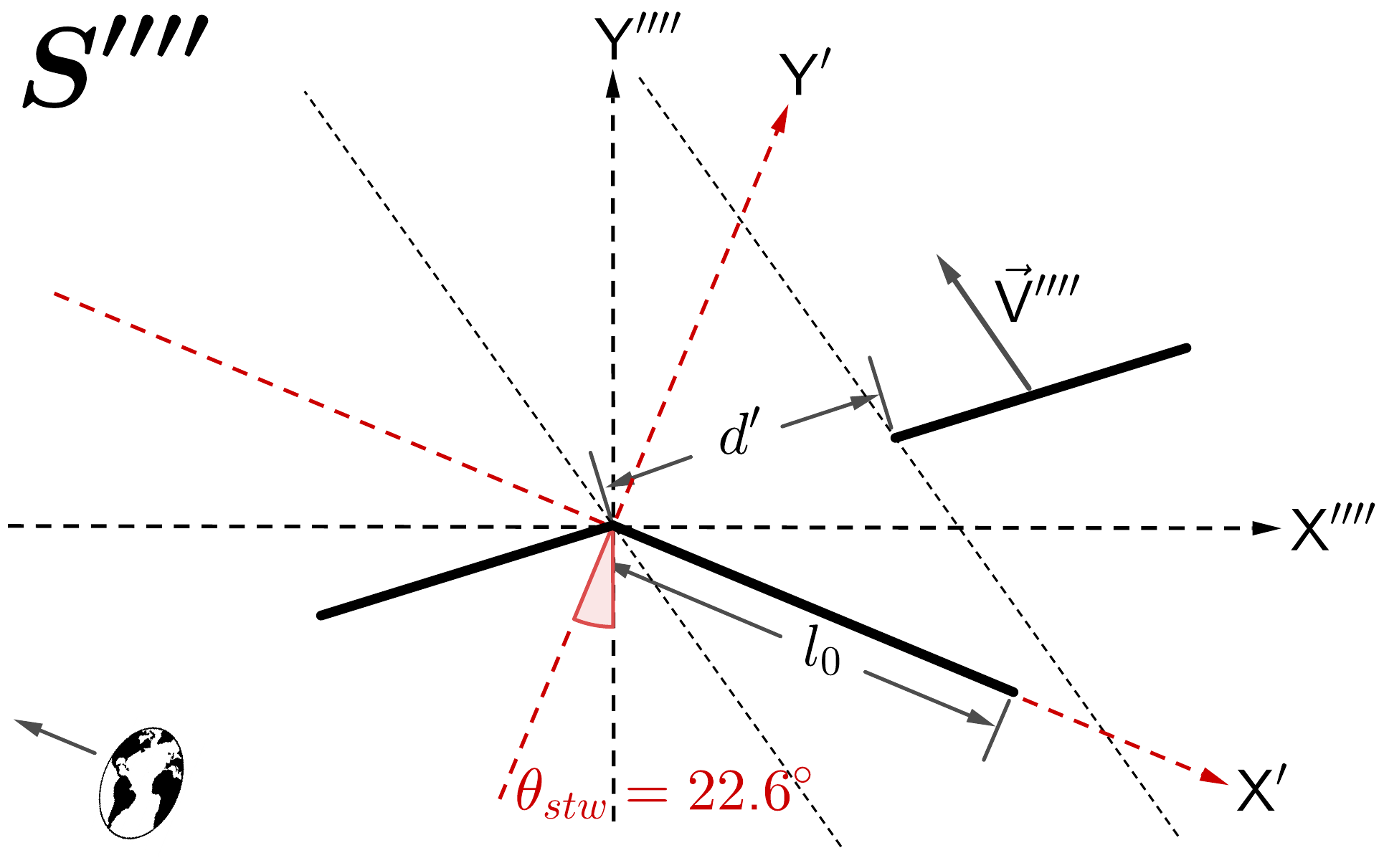}}}
    
    \caption{\sl Shows the STW-rotation when (a) transforming from $S'$ directly to the rest frame of the slit causing a new observer’s reference $S'''$ which is STW-rotated $22.6 \degree$ clockwise relative to $S''$, or the complimentary (b) $S''$ to the rest frame of the rod causing a new observer’s reference $S''''$ rotated $22.6 \degree$ counterclockwise relative to $S'$. The globe in the lower left corner shows the relative direction of motion of frame $S$.}
    \label{fig4}
\end{figure}

The STW-rotation between the rod and the slit is not obvious in Figs. \ref{fig2}b and \ref{fig2}c because it is mixed with rotation caused by length contraction. In order to see that the STW-rotation is a real rather than apparent effect, we can study the direct transformation between frames $S'$ and $S''$, as shown in Fig. \ref{fig4}. A direct transformation from $S'$ to the rest frame of the slit is displayed in Fig. \ref{fig4}a, which is the STW-rotated version of Fig. \ref{fig2}c. This system has been labeled $S'''$, and the STW-rotated axes of $S''$ have been added to the figure. To emphasize the complimentary nature of relativity, the direct transformation from $S''$ to the rest frame of the rod is displayed in Fig. 4b, which is a rotated version of Fig. \ref{fig2}b. Similarly, this system has been labeled $S''''$, and is shown to be rotated by the same STW-angle but in the opposite direction of the system in Fig. \ref{fig4}a.

-

\begin{figure}[h!]
    \centering
    \fbox{\subfigure[]{\includegraphics[width=0.30\textwidth]{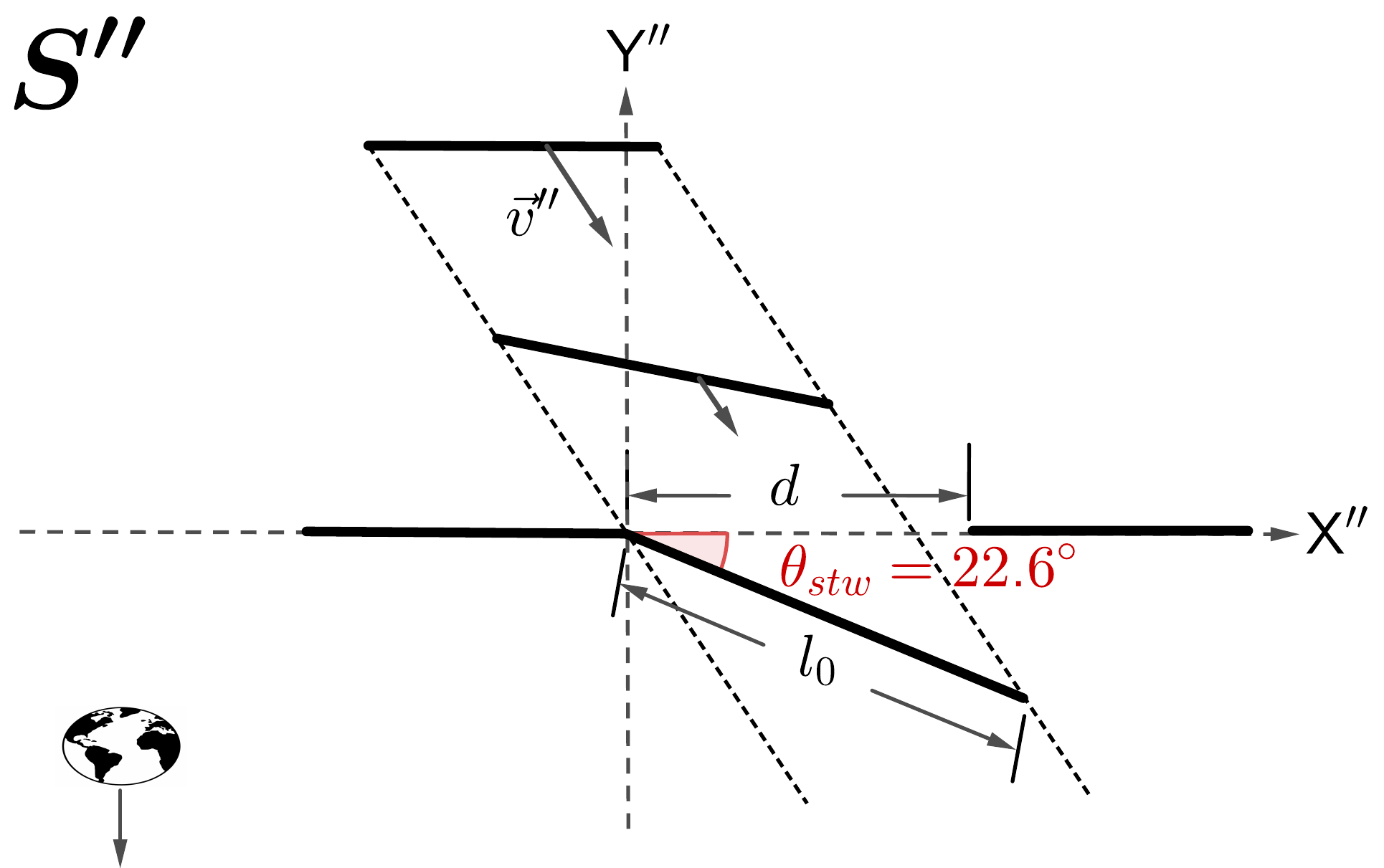}}}
       
   \fbox{ \subfigure[]{\includegraphics[width=0.30\textwidth]{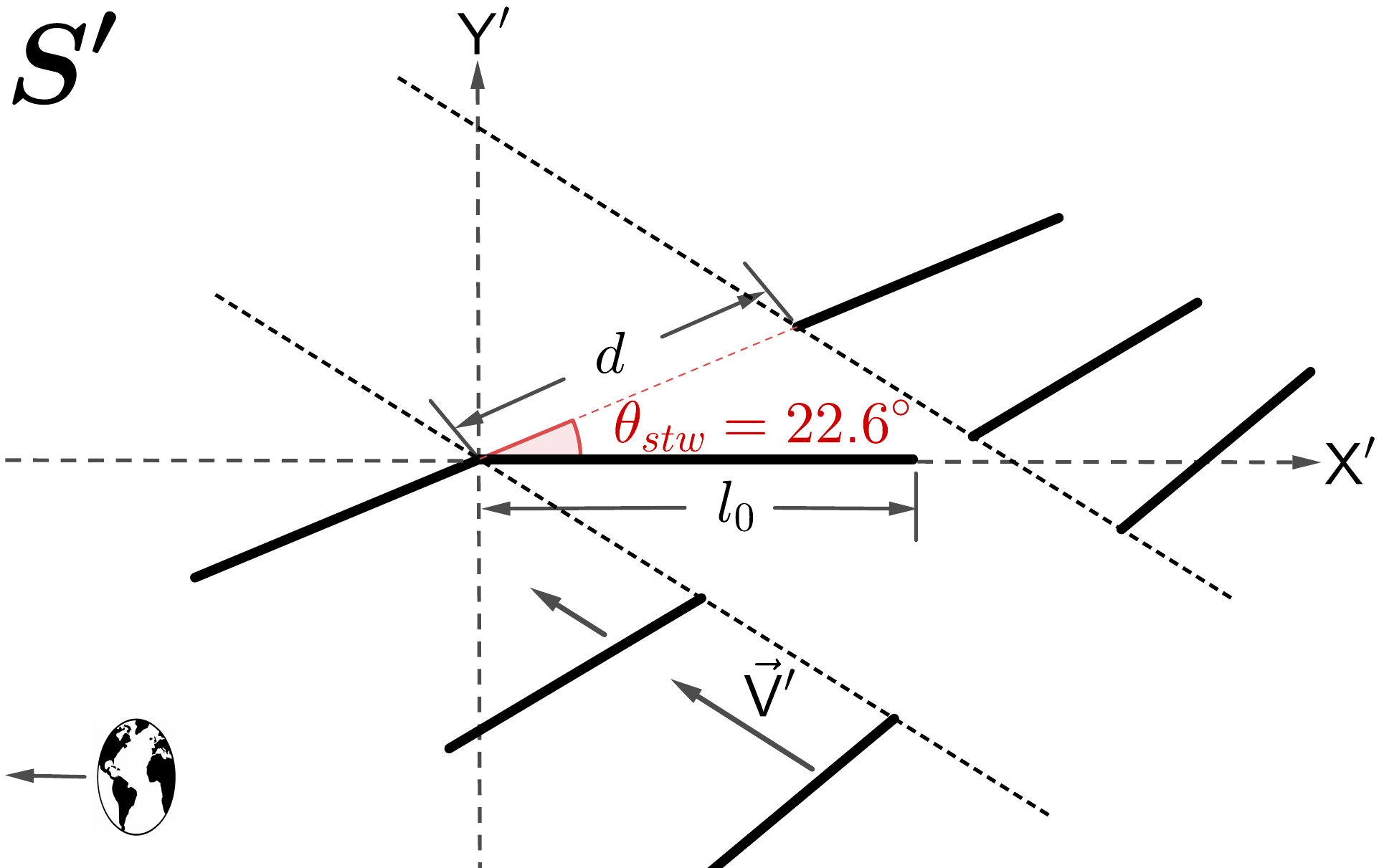}}}
    
    \caption{\sl Two new final states of the initial setup defined in Fig. 2 to better exemplify the real effect of the STW-rotation. Both new states bring the sheet and rod to rest relative to each other via two different routes. (a) Starts in the $S''$ rest frame of the sheet per Fig. \ref{fig2}c, and reduces the velocity of the rod directly along $\vec{v''}$ to a stop as points 1 and 3 coincide in order to show the rod is now STW-rotated relative to the slit by $22.6 \degree$. The complimentary situation in (b) starts in the $S'$ rest frame of the rod shown in Fig. \ref{fig2}b and reduces the velocity of the sheet directly along $\vec{V'}$ to a stop as points 1 and 3 coincide in order to show the sheet is now STW-rotated relative to the rod. The globe in the lower left corner shows the relative direction of motion of frame $S$.}
    \label{fig5}
\end{figure}

Another way to show the STW-rotation directly is by introducing a new final system where a boost along $\vec{v''}$ is performed in the $S''$ system, to bring the rod to rest. Because the boost is parallel to the direction of relative motion, there is no change to the STW-angle. The result is displayed in Fig. \ref{fig5}a. As the velocity of the rod stops, the rotation caused by length contraction is removed so that only the STW-rotation of $22.6 \degree$ is left. To be comprehensive and keep with the theme of showing the complimentary conditions, we also perform a boost along $\vec{V'}$ in the $S'$ system, to bring the sheet to rest Again the boost is in the direction of motion and the result is displayed in Fig. \ref{fig5}b. The rotation caused by length contraction on the sheet is removed and the STW-rotation of $22.6 \degree$ is visible and in the opposing (complimentary) direction in Fig. \ref{fig5}a.

\section{Conclusion}

In regard to the treatment by E. Marx. \cite{Marx} (1967), we argued that in some of the figures the direction of motion is misrepresented, and that the dimensions of the slit are distorted. The set of comparable properly scaled figures in Fig. \ref{fig2} were included to clarify this potential point of confusion, and to illustrate the relevant, and too-often neglected phenomenon of the STW-rotation. In regard to the treatment by Iyer and Prabhu \cite{Prabhu} (2006), we argued that their conclusion is incorrect because they fail to recognize the rotation in Shaw’s original experiment \cite{Shaw} as a relativistic effect. Because they only analyzed the system for the effect of a single, linear velocity transformation between two frames $S'$ and $S''$ and neglected the effect of the non-collinear boosts relative to $S$ that decide the outcome of the original experiment, their conclusions only apply to Marx's experiment I. While the STW or Wigner rotation is not directly evident in the change in orientations of the rod and slit between the different reference frames of Fig. \ref{fig2}, the rotation indirectly plays a key role in both the solution to the paradox, and the explanation for the change in relative orientation of the components. Finally, Figs. 4 and 5 were introduced to illustrate the STW-rotations of the rod, slit, and observer coordinate systems to clarify this point.

\section*{Declarations}

\textbf{Ethics Approval} Not applicable.
\\
\textbf{Conflicts of Interest} The author has no relevant financial or non-financial interests to disclose.

\appendix

\section{Solution to the rod and slit paradox}

To assist in understanding the resolution to the apparent paradox of the rod and slit, as proposed by Shaw \cite{Shaw}, an explanation defining the world lines of the rod and sheet with slit, is presented in this Appendix. The solution closely follows Marx \cite{Marx} and Ferraro \cite{Ferraro_book}. 

Assuming that the initial conditions are suitably defined as shown in frame $S$ of Fig. \ref{fig2}a, and that the contracted rod length $l$ is slightly smaller than the slit of the sheet, then the rod can pass through the slit. The apparent paradox of the experiment arises because the length of the rod may be longer than the contracted length of the slit in the complimentary frame defined as $S'$ in Fig. \ref{fig2}b. 
The seemingly paradoxical nature of the thought-experiment is resolved when the change in the orientation of the slit is considered. To consider the change in orientation of the sheet and slit between $S$ and $S'$, compare the simultaneous positions of points $3$ and $4$ on the sheet. World lines of points $1$, $2$, $3$, and $4$ are described by

\begin{equation*}
\begin{aligned}[c]
x_{1} & =vt \\
x_{2} & =vt+l \\
x_{3} & =0 \\
x_{4} & =d
\end{aligned}
\qquad
\begin{aligned}[c]
y_{1} & =0 \\
y_{2} & =0 \\
y_{3} & =Vt \\
y_{4} & =Vt
\end{aligned}
\qquad
\begin{aligned}[c]
t_{1} & =t \\
t_{2} & =t \\
t_{3} & =t \\
t_{4} & =t .
\end{aligned}
\qquad
\begin{aligned}[c]
(A.1) \\
(A.2) \\
(A.3) \\
(A.4)
\end{aligned}
\end{equation*}

The common parameter is chosen as coordinate time, $t$, and the contracted length of the rod is $l = l_0/\gamma_v$ where $\beta_v = v/c$  and $\gamma_v = (1 - \beta_v^2)^{-1/2}$. Then, to perform a pure Lorentz transformation to the frame of the rod moving in the positive x-direction with velocity $v$, use the linear transform between four-vectors in matrix form 

\begin{equation*}
\begin{pmatrix}
ct' \\
x' \\
y' \\
z' \\
\end{pmatrix}
=
\begin{pmatrix}
\gamma_v & -\gamma_v \beta_v & 0 & 0 \\
- \gamma_v \beta_v & \gamma_v & 0 & 0 \\
0 & 0 & 1 & 0 \\
0 & 0 & 0 & 1 \\
\end{pmatrix}
\begin{pmatrix}
ct \\
x \\
y \\
z \\
\end{pmatrix}
\tag{A.5}
\label{x_transformation}
\end{equation*}

In terms of the new time parameter, $t'$, the world line equations become

\begin{equation*}
\begin{aligned}[c]
x_{1}' & =0 \\
x_{2}' & =\gamma_v l \\
x_{3}' & =-vt' \\
x_{4}' & =-vt'+ d/\gamma_v
\end{aligned}
\quad
\begin{aligned}[c]
y_{1}' & =0 \\
y_{2}' & =0 \\
y_{3}' & =Vt'/\gamma_v \\
y_{4}' & =Vt'/\gamma_v + \beta_v \beta_{\scriptscriptstyle V} d
\end{aligned}
\quad
\begin{aligned}[c]
t_{1}' & =t' \\
t_{2}' & =t' \\
t_{3}' & =t' \\
t_{4}' & =t'
\end{aligned}
\quad
\begin{aligned}[c]
(A.6) \\
(A.7) \\
(A.8) \\
(A.9)
\end{aligned}
\end{equation*}

\noindent where $\beta_{\scriptscriptstyle V} = V/c$. The spacetime trajectories corresponding to Eq. (A.6) through (A.9) are shown in Fig. \ref{fig2}(b), where the rod and the sheet with a slit are at a set time $t' = 0$. At this time point $1$ coincides with point $3$. The length of the rod is then $l_0$, and the length of the slit is found to be 

\begin{equation*}
d' = (1-\beta_v^2 (1-\beta_{\scriptscriptstyle V}^2))^{1/2}d < d < l_0 .
\tag{A.10}
\end{equation*}

Because the sheet is slanted with respect to the rod, it will still pass through the slit. This can be checked by finding the event where the far end of the slit meets the $x'$ axis. By choosing $t'_4 = - \beta_v \gamma_v d/c $ we get $y'_4=0$ and

\begin{equation*}
x'_4 = \gamma_v d > \gamma_v l = l_0 .
\tag{A.11}
\end{equation*}

The orientation of the slit in $S'$ differs from the one in $S$ since  $y'_4  \neq y'_3$ in $S'$ (as can be seen in Fig. \ref{fig2}b). The angle between the sheet and the $x'$ axis is given by

\begin{equation*}
tan(\Psi) = \frac{y'_4 - y'_3}{x'_4 - x'_3} = \gamma_v \frac{vV}{c^2} = \gamma_v \beta_v \beta_{\scriptscriptstyle V} .
\tag{A.12}
\end{equation*}

Note that this angle $\Psi$ is not the STW-rotation angle since the sheet is length-contracted along the velocity $\vec{V'}$ in the $S'$ frame.

To obtain the coordinates in the $S''$ frame as displayed in Fig. \ref{fig2}c, a Lorentz transformation from $S$ in the positive y-direction with velocity $V$ is performed. In this case the linear transform between four-vectors in matrix form is given

\begin{equation*}
\begin{pmatrix}
ct'' \\
x'' \\
y'' \\
z'' \\
\end{pmatrix}
=
\begin{pmatrix}
\gamma_{\scriptscriptstyle V} & 0 & -\gamma_{\scriptscriptstyle V} \beta_{\scriptscriptstyle V} & 0 \\
0 & 1 & 0 & 0 \\
- \gamma_{\scriptscriptstyle V} \beta_{\scriptscriptstyle V} & 0 & \gamma_{\scriptscriptstyle V}  & 0 \\
0 & 0 & 0 & 1 \\
\end{pmatrix}
\begin{pmatrix}
ct \\
x \\
y \\
z \\
\end{pmatrix} .
\tag{A.13}
\label{y_transformation}
\end{equation*}

Where $\gamma_{\scriptscriptstyle V} = (1 - \beta_{\scriptscriptstyle V}^2)^{-1/2}$. Based on a new time parameter $t''$, the world line equations then become

\begin{equation*}
\begin{aligned}[c]
x_{1}'' & =vt''/ \gamma_{\scriptscriptstyle V} \\
x_{2}'' & =vt''/\gamma_{\scriptscriptstyle V} + l \\
x_{3}'' & =0 \\
x_{4}'' & =d
\end{aligned}
\qquad
\begin{aligned}[c]
y_{1}'' & =-Vt'' \\
y_{2}'' & =-Vt'' \\
y_{3}'' & =0 \\
y_{4}'' & =0
\end{aligned}
\qquad
\begin{aligned}[c]
t_{1}'' & =t'' \\
t_{2}'' & =t'' \\
t_{3}'' & =t'' \\
t_{4}'' & =t'' .
\end{aligned}
\qquad
\begin{aligned}[c]
(A.14) \\
(A.15) \\
(A.16) \\
(A.17) 
\end{aligned}
\end{equation*}

These world line trajectories in Fig. \ref{fig2}c clearly show the rod is still contracted to length $l$ and remains parallel to the sheet and to the x-axis. Since the slit is of length $d$ when at rest and $l < d$ the rod will pass through the slit in this frame as well.

\end{document}